\documentclass[aps,amsmath,amssymb,amsfonts,twocolumn,showpacs]{revtex4}
\usepackage{graphicx}
\usepackage{subfigure}
\usepackage{amsmath,amssymb,amsfonts,mathrsfs}
\usepackage[dvips]{color}

\newcommand{\f}[1]{\textrm{#1}}

\newcommand{\x}{{\mathbf r}}

\begin{document}

\title{Simple model of Feshbach resonance in the strong-coupling regime}

\author{T. Wasak, M. Krych, Z. Idziaszek and M. Trippenbach}
\affiliation{Faculty of Physics, University of
Warsaw, ul.  Ho\.{z}a 69, PL--00--681 Warszawa, Poland}

\author{Y. Avishai}
\affiliation{Department of Physics and the Ilse Katz Center
for Nano-Science, Ben-Gurion University, Beer-Sheva 84105, Israel}

\author{Y. B. Band}
\affiliation{Department of Chemistry, Department of Physics and
Department of Electro-Optics, and the Ilse Katz Center for
Nano-Science, Ben-Gurion University, Beer-Sheva 84105, Israel}

\begin{abstract}
We use the dressed potentials obtained in the adiabatic representation
of two coupled channels to calculate $s$-wave Feshbach resonances in a
3D spherically symmetric potential with an open channel interacting
with a closed channel.  Analytic expressions for the $s$-wave
scattering length $a$ and number of resonances are obtained for a
piecewise constant model with a piecewise constant interaction of the
open and closed channels near the origin.  We show analytically and
numerically that, for strong enough coupling strength, Feshbach
resonances can exist even when the closed channel does {\em not} have
a bound state.
\end{abstract}

\maketitle


\section{Introduction}

Resonance scattering phenomena are ubiquitous; they have been observed
in many different physical scenarios, including electronic transport
and optical spectra in semiconductors \cite{Faist, BarAd, Kroner},
collisions of ultracold atoms \cite{Chin_10a} and creation of
ultracold molecules \cite{Kohler}.  Herman Feshbach \cite{Feshbach}
and Ugo Fano \cite{Fano} developed theoretical methods to treat
resonance phenomena that arise from the coupling of a discrete state
to the continuum.  This type of resonance scattering was first modeled
in the context of the autoionization of atoms \cite{Fano}.  The term
Feshbach resonance (FR), sometimes also called Fano-Feshbach
resonance, is now most often used to describe scattering processes
wherein a bound state on a ``closed channel'' strongly affects the
scattering on an open scattering channel which is weakly coupled to
the bound state; flux entering the open channel gets temporarily
caught in the closed channel bound state, before eventually leaking
back to the open channel.  The use of FR scattering to affect the low
energy collisions of atoms has become an important experimental tool
for controlling the properties of low temperature atomic and molecular
gases \cite{Chin_10a}.  Thus, the magnitude and sign of the low-energy
atomic interactions can be varied by coupling free particles to a
molecular state \cite{Moerdijk,Timmermans}, thereby tuning the
properties of the many-body gas.  Mostly, magnetic field induced
Feshbach scattering resonances have been employed experimentally.  In
addition, FRs can also be controlled optically
\cite{Bauer_09,Blatt_11} or with the help of electric fields
\cite{Li}.  For example, scattering resonances arise under the
influence of laser light tuned near a photoassociation resonance,
where free atom pairs are coupled to an excited molecular state
\cite{Jones_06}.

In this work we introduce a FR model for 3D scattering from centrally
symmetric piecewise constant open and closed channel potentials with a
constant interaction over the inner part of the potentials. The model
admits an analytic solution that allows for a full understanding of
the various aspects of FR scattering that have heretofore not been explored.
Specifically, we show that a new regime of FRs exists wherein a closed
channel with a {\it repulsive potential} {\footnote{In our case a repulsive potential is a square well potential with a positive short range part.}} can give rise to FRs when
the interaction coupling strength is sufficiently strong; in this
regime, FRs can exist even when the closed channel does not have a bound
state.  The present analysis complements the tight-binding model and
the formal scattering theory results published in Ref.~\cite{ABT_13}, {square well model presented in Ref.~\cite{Lange_09} and atomic waveguide model from Ref.~\cite{Saeidian_12}},
and demonstrates that a FR with a closed channel that does not have a
bound state is not an artifact of the tight-binding model.

The outline of this manuscript is as follows.  In Sec.~\ref{basic}
the basic equations defining the FR physics are presented. In
Sec.~\ref{Sec:piecewise} we specify the model, define the
piecewise-constant coupling model and introduce the dimensionless units
used in the manuscript.
Section~\ref{sec:cc} solves the coupled channel Schr\"odinger equation
for bound states.  The analytic formula for the scattering length is
derived in Sec.~\ref{sec:scatt}.  Section~\ref{sec:adiab} presents the
adiabatic representation and the bound states that appear in the
``dressed'' potentials of this representation (the dressing is due to
the interaction between channels).  In Sec.~\ref{sec:nonpert} we
present a detailed analysis of the resonances in the {strong coupling}
regime.  Finally, a summary and conclusion are provided in
Sec.~\ref{Sec:Summary}.

\section{Basic equations defining Feshbach resonance physics} \label{basic}

We consider two coupled channels, the open channel $o$ and the closed
channel $c$.  The wave functions $\psi_o$ and $\psi_c$ satisfy
Schr\"odinger equation,
\begin{eqnarray}
\label{schroedinger_general}
\left(-\frac{\hbar^2}{2m}\nabla^2 + U_o(\x)\right) \psi_o(\x) +
U_{oc}(\x)\psi_c(\x) &=& E \psi_o(\x) , \\
\left(-\frac{\hbar^2}{2m}\nabla^2 + U_c(\x)\right) \psi_c(\x) +
U_{co}(\x)\psi_o(\x) &=& E \psi_c(\x) .
\end{eqnarray}
We assume that the open channel potential $U_o(\x)$
vanishes as $|\x| \to \infty$, i.e., the open channel
asymptote defines the zero of potential energy.  The closed channel
potential $U_c(\x)$ tends asymptotically to a positive constant
$\nu > 0$.  We further assume that the coupling potential
$V_{oc}(\x)$ is real.  Figure~\ref{fig:potentialadiab}(a) shows a
schematic illustration of the open and closed potentials, $U_o$ and
$U_c$, which will be fully specified in Sec.~\ref{Sec:piecewise}.

\section{Piecewise-constant Coupling}
\label{Sec:piecewise}

\begin{figure}
\includegraphics[width=0.5\textwidth]{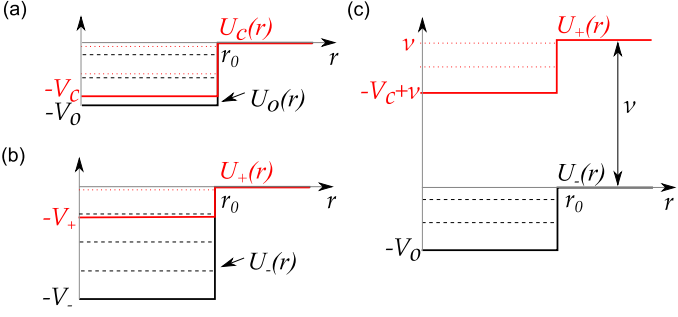}
\caption{(Color online) Dressed potential picture of coupled channel
model for piecewise constant potential.  (a) The level structure of
original potentials $U(r)$.  Open channel $U_o(r)$ and closed channel
$U_c(r)$ (coupling $\tau=0$ and detuning $\nu=0$).  (b) Dressed
potentials $U_+(r)$ and $U_-(r)$ for non-vanishing coupling, $\tau>0$, and
asymptotic potential energy difference $\nu=0$.  (c) Same as (b) but
for $\nu \gg 0$.  In all panels, dashed lines denote bound states of
the $U_o(r)$ and $U_-(r)$ potentials and dotted lines correspond to $U_c(r)$
and $U_+(r)$. All length variables are expressed in units of $r_0$ and all energies are given in units of $\hbar^2/(2mr_0^2)$.
}
\label{fig:potentialadiab}
\end{figure}

We use a square well model with piecewise constant potentials of the
form (see Fig.~\ref{fig:potentialadiab}),
\begin{equation}  \label{eqn:potential_Uo}
U_o(r) = \left\{
\begin{array}{ll}
0 & \textrm{if $r>r_0$}\\
-V_o & \textrm{if $r<r_0$}
\end{array} \right. ,
\end{equation}
\begin{equation}  \label{eqn:potential_Uc}
U_c(r) = \left\{
\begin{array}{ll}
\nu & \textrm{if $r>r_0$}\\
-V_c + \nu & \textrm{if $r<r_0$}
\end{array} \right. ,
\end{equation}
with $\nu \geqslant 0$, and that the open and closed channels
are coupled inside the well (see, e.g., Ref.~\cite{Chin_10}),
\begin{equation}  \label{eqn:piecewise_coupling}
    U_{oc}(r) = g\, \theta(1-|r_0|),
\end{equation}
where $\theta$ is the step function and the coupling interaction
strength $g$ is real.  The Schr{\"o}dinger equation for the $s$-wave
radial functions $u_{o,c}(r)=r \psi_{o,c}(r)$ with energy $E$ is
\begin{equation}
-\frac{\hbar^2}{2m}u''_o(\rho) + (U_o(\rho)-E)u_o(\rho) +
U_{oc}(\rho)u_c(\rho) = 0 ,
\end{equation}
\begin{equation}
-\frac{\hbar^2}{2m}u''_c(\rho) + (U_c(\rho)-E)u_c(\rho) +
U_{oc}(\rho)u_o(\rho) = 0 .
\end{equation}

The energy scales in the model can be related to the wavenumbers.
Below we use the dimensionless wavenumbers related to the energies by
the energy-momentum relations: $\kappa = \sqrt{2mr_0^2 E}/\hbar$,
$\kappa_{o,c} = \sqrt{2m r_0^2 V_{o,c}}/\hbar$, and $\kappa_{\nu} =
\sqrt{2mr_0^2 \nu}/\hbar$, where $\nu$ is the asymptotic potential
energy difference and $m$ is a reduced mass.  Moreover, all length
variables will hereafter be expressed in units of $r_0$ and all
energies will be given in units of $\hbar^2/(2mr_0^2)$.  We also use a
dimensionless coupling strength $\tau = (2mr_0^2g)/\hbar^2$,
dimensionless scattering lengths, $\alpha = a/r_0$ and $\alpha_{bg} =
a_{bg}/r_0$, and a dimensionless radial coordinate $\rho = r / r_0$.

\section{Coupled channel Schr{\"o}dinger equation: Bound states}
\label{sec:cc}

The Schr{\"o}dinger equation for the square well model introduced in
Sec.~\ref{Sec:piecewise} can be solved analytically \cite{Chin_10}.
For $\rho > 1$, the inter-channel coupling is zero, and the
Schr{\"o}dinger equations  for bound states can be written as
\begin{eqnarray}\label{schrout}
  -\partial_\rho^2 u_o = -\kappa^2 u_o,\\
  (-\partial_\rho^2 + \kappa_{\nu}^2)u_c = -\kappa^2 u_c,
\end{eqnarray}
and their solutions are,
\begin{eqnarray}
  u_o(\rho) &=& A_o e^{-\kappa \rho} ,  \nonumber \\
  u_c(\rho) &=& A_c e^{-\sqrt{\kappa^2+\kappa_{\nu}^2} \rho} .
\end{eqnarray}
In the region where the channels are coupled, $\rho < 1$, the equations are,
\begin{eqnarray} \label{schr}
(-\partial_\rho^2 - \kappa_o^2)u_o + \tau u_c &=& -\kappa^2 u_o, \nonumber \\
(-\partial_\rho^2 - \kappa_c^2+\kappa_{\nu}^2)u_c + \tau u_o &=& -\kappa^2 u_c.
\end{eqnarray}
To solve Eqs.~(\ref{schr}) we introduce the superposition states
\begin{equation} \label{vVSu}
  v_o = c \, u_o - s \, u_c, \quad
  v_c = s \, u_o + c \, u_c, \nonumber
\end{equation}
where $c \equiv \cos \theta$, $s \equiv \sin \theta$, and
\begin{equation}
    \tan(2\theta) = -\frac{2\tau}{(\kappa_o^2-\kappa_c^2 +
    \kappa_{\nu}^2)},
\end{equation}
and we obtain the set of uncoupled equations,
\begin{eqnarray}
\left( \partial_\rho^2 + c^2 \kappa_o^2 + s^2
(\kappa_c^2 -\kappa_{\nu}^2)- \kappa^2
- 2\tau sc\right) v_o &=& 0 , \label{diffeq1} \\
\left( \partial_\rho^2 + s^2 \kappa_o^2 + c^2
(\kappa_c^2 -\kappa_{\nu}^2)- \kappa^2
+ 2\tau sc\right) v_c &=& 0.\label{diffeq2}
\end{eqnarray}
Note that the mixing angle $\theta$ does not depend on energy; it only
depends on the potentials and coupling strength $\tau$.  We define two
auxiliary quantities,
\begin{eqnarray}
\xi_1 &=& \sqrt{c^2 (\kappa_o^2 - \kappa^2) + s^2
(\kappa_c^2 - \kappa^2 -\kappa_{\nu}^2)
- 2\tau sc} , \nonumber \\
\xi_2 &=& \sqrt{s^2 (\kappa_o^2 - \kappa^2) + c^2
(\kappa_c^2 - \kappa^2 -\kappa_{\nu}^2)
+ 2\tau sc} \nonumber
\end{eqnarray}
and notice that they can be either real or pure imaginary.  Using the
above definitions, the solution of the differential equations
(\ref{diffeq1}) and (\ref{diffeq2}) are,
\begin{equation}
    v_o(\rho) = C_o \, \sin(\xi_1 \rho) , \quad v_c(\rho)= C_c \,
    \sin(\xi_2 \rho) ,
\end{equation}
which satisfy boundary conditions, $v_o(0)=v_c(0)=0$.  In order to
obtain the energy of the bound states and the scattering length of the
coupled system, we consider the following form of the wave functions
$u_o(\rho)$ and $u_c(\rho)$:
\begin{eqnarray}
u_o(\rho) &=& c \, C_o \sin(\xi_1 \rho) + s \, C_c \sin(\xi_2 \rho), \quad
\f{ for } \rho < 1  \nonumber \\
 \quad u_o(\rho) &=& A_o e^{-\kappa \rho}, \quad \f{
for } \rho > 1 \nonumber \\
u_c(\rho) &=& -s \, C_o \sin(\xi_1 \rho) + c \, C_c \sin(\xi_2 \rho), \quad
\f{ for } \rho < 1 \nonumber \\
 \quad u_c(\rho) &=& A_c
e^{-\sqrt{\kappa^2+\kappa_{\nu}^2} \rho}, \quad \f{ for } \rho > 1.
\end{eqnarray}
Note that in order to determine the scattering length, one has to take
wavefunction for the zero energy, and for $\rho>1$, $u_o(r) =
A_o(\rho-\alpha)$.

To determine the bound states, we match the logarithmic derivatives of
$u_o$ and $u_c$ at $\rho = 1$, assuming $\kappa>0$, and thereby obtain
the two conditions,
\begin{eqnarray}
\frac{c \, C_o \xi_1 \cos(\xi_1) + s C_c \xi_2 \cos(\xi_2)}
{c \, C_o \sin(\xi_1) + s C_c \sin(\xi_2) } &=& -\kappa,\\
\frac{-s \, C_o \xi_1 \cos(\xi_1) + c \, C_c \xi_2 \cos(\xi_2)}
{-s \, C_o \sin(\xi_1) + c \, C_c \sin(\xi_2) } &=&
-\sqrt{\kappa^2+\kappa_{\nu}^2}.
\end{eqnarray}
We can extract $C_c/C_o = s \zeta_b/c$ from the first of these
equations.  Defining $\zeta_b$, we find,
\begin{equation}
    \zeta_b = - \frac{ \xi_1 \cos(\xi_1) + \kappa \sin(\xi_1) }{ \xi_2
    \cos(\xi_2) + \kappa \sin(\xi_2) },
\end{equation}
and then inserting this ratio into the second equation to obtain an
expression for energy $E = -\hbar^2\kappa^2/(2m r_0^2)$, we obtain
\begin{equation} \label{pwcbound}
    \sqrt{\kappa^2 + \kappa_{\nu}^2} = \frac{ s^2 \xi_1 \cos(\xi_1) +
    c^2 \zeta_b \xi_2 \cos(\xi_2) }{ s^2 \sin(\xi_1) + c^2\zeta_b
    \sin(\xi_2) }.
\end{equation}
To obtain the values of the energies of the eigenstates of the coupled
system, Eq.~(\ref{pwcbound}) needs to solved for $\kappa$.

\section{The Scattering Length}  \label{sec:scatt}

To determine the scattering length, we match the logarithmic
derivatives of the zero energy ($\kappa=0$) wavefunctions $u_o$ and
$u_c$ at $\rho = 1$, and find,
\begin{eqnarray}
\left. \frac{u'_o(1^{-}) }{u_o(1^{-})}\right|_{\kappa=0} &=&
\frac{1}{1-\alpha} \label{pwcsl1},\\
\left. \frac{u'_c(1^{-}) }{u_c(1^{-})}\right|_{\kappa=0} &=&
-\kappa_{\nu}. \label{pwcsl2}
\end{eqnarray}
From Eq.~(\ref{pwcsl2}), we obtain $\zeta_a$, defined by relation
$C_c/C_o = s \zeta_a / c$:
\begin{equation}
\zeta_a = \frac{\xi_1
    \cos(\xi_1) + \sqrt{\kappa^2+\kappa_{\nu}^2} \sin(\xi_1) }{ \xi_2
    \cos(\xi_2) + \sqrt{\kappa^2+\kappa_{\nu}^2} \sin(\xi_2) }.
\end{equation}
Using Eq.~(\ref{pwcsl1}), an expression for the scattering length,
expressed in terms of $\zeta_a$, is obtained:
\begin{equation}  \label{Eq:apha}
    \alpha = \left.  1-\frac{ c^2 \sin(\xi_1) + s^2\zeta_a\sin(\xi_2)
    }{ c^2 \xi_1 \cos(\xi_1) + s^2 \zeta_a \xi_2 \cos(\xi_2)
    }\right|_{\kappa=0}.
\end{equation}

\section{Adiabatic representation} \label{sec:adiab}

We determine the number of FRs for the whole range of asymptotic
potential energy differences between the coupled channels, $\nu$.  We
use the adiabatic representation in which the Hamiltonian describing
the problem [Eqs.~\eqref{schrout} and \eqref{schr}] is transformed
unitarily to make the potential matrix diagonal.  For piecewise
constant interaction potentials, the kinetic energy operator remains
diagonal.  The eigenvalues of the potential energy matrix, denoted by
$U_{\pm}$, are such that
\begin{equation}
0=\textrm{det}\!
\begin{pmatrix}
U_o - U_{\pm}& U_{oc}\\
U_{oc} & U_c- U_{\pm},
\end{pmatrix}
\end{equation}
The eigenvalues for $r<r_0$ are $-V_\pm$, where
\begin{equation}  \label{veff}
    V_\pm = \frac{V_o + V_c-\nu}{2} \mp
    \sqrt{\left(\frac{-V_o+V_c-\nu}{2}\right)^2+V_{oc}^2} .
\end{equation}

In case of the piecewise constant potential with piecewise constant
coupling, the adiabatic representation yields an exact solution
everywhere.  Notice that
adiabatic transformation matrix is constant and commutes with the
kinetic energy operator.  Equation~\eqref{veff} describes two
effective potentials.  The stronger the coupling between the channels,
the larger is the change in the well depth.  The higher energy potential is
shifted upwards by the same amount as the lower one is shifted
downwards.  This can result in the emergence of new bound states in
the lower channel and their vanishing in the upper one (cf.,
Fig.~\ref{fig:potentialadiab}).

\begin{figure}[!ht]
\centering
\centering\subfigure[]{\includegraphics[width=0.45\textwidth]{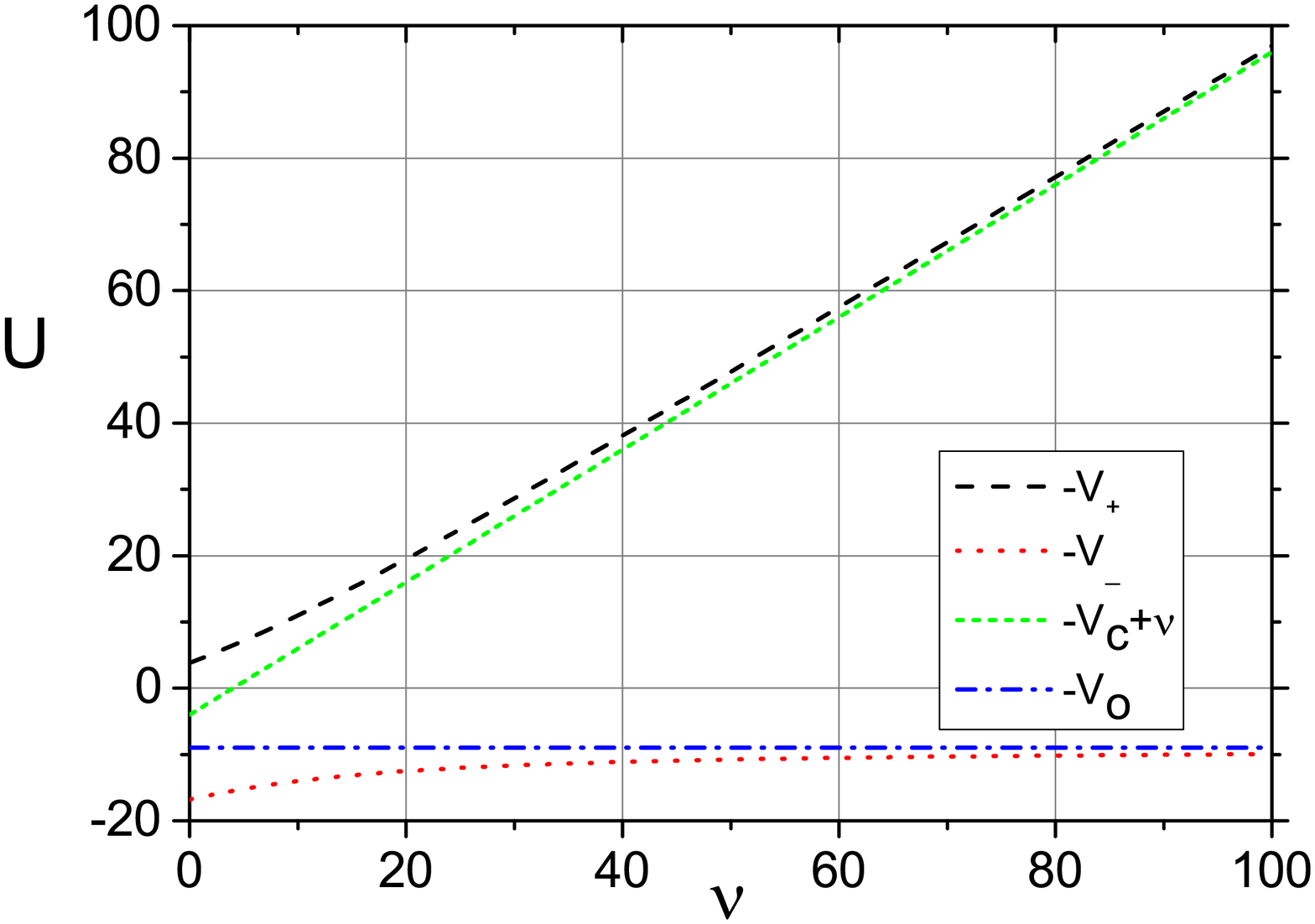}}
\centering\subfigure[]{\includegraphics[width=0.45\textwidth]{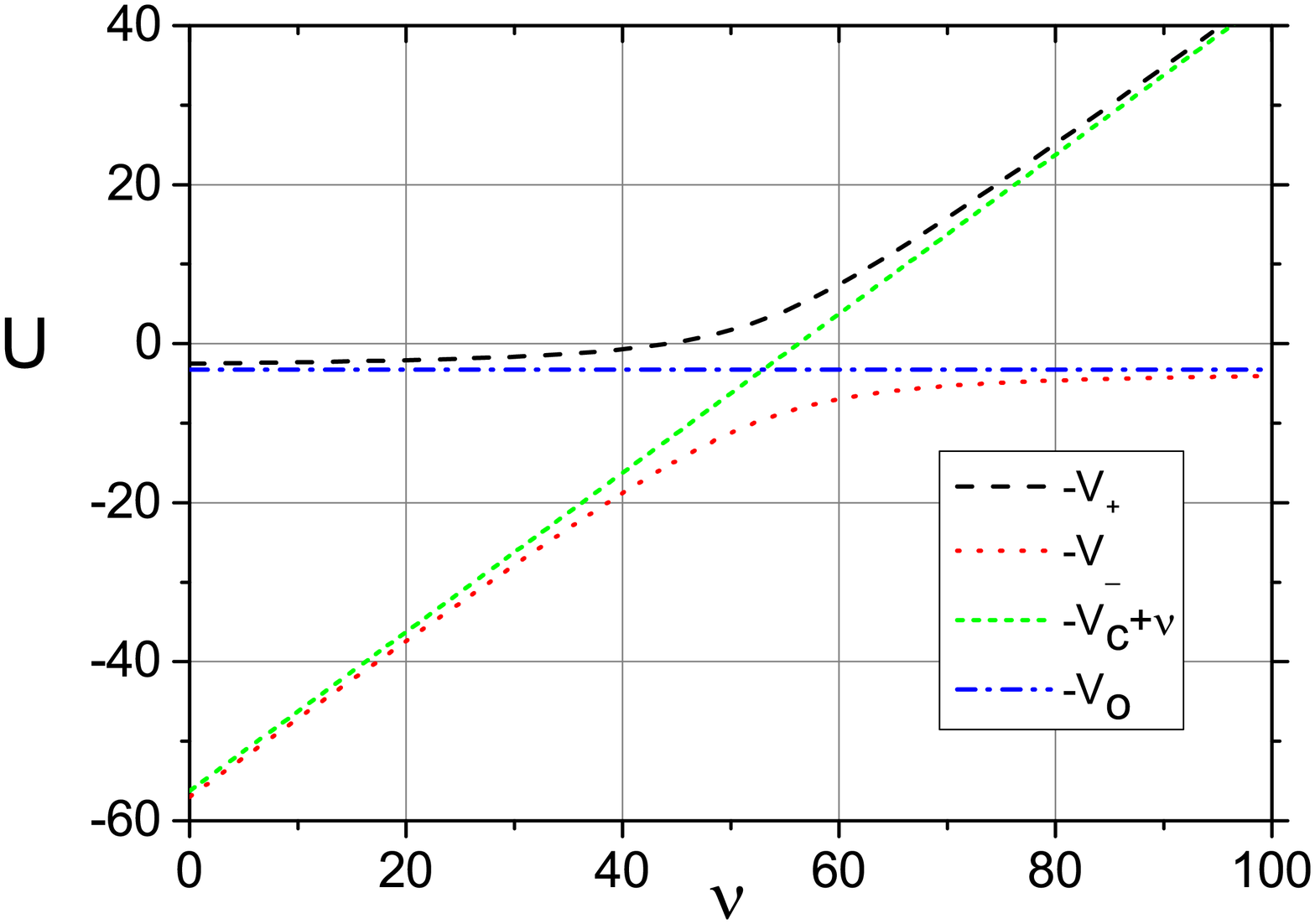}}
\caption{(Color online) The depth of the dressed potentials versus the
detuning $\nu$, depending on the relative depth of $V_o$ and $V_c$.
(a) $V_c < V_o$, and in particular $\kappa_o=3$, $\kappa_c=2$,
$\tau=10$ (b) $V_o < V_c$ ($\kappa_o=1.8$, $\kappa_c=7.5$,
$\tau=6.3$), where the avoided crossing is clearly seen. All energies are given in units of $\hbar^2/(2mr_0^2)$.}
\label{structure1}
\end{figure}

It is important to analyze the depth of the potentials $V_+$ and $V_-$
in order to establish their bound state structure.  From
Eq.~(\ref{veff}) we conclude that they depend both on the coupling
$V_{oc}$ and the detuning $\nu$.  In Fig.~\ref{structure1} we plot
both $V_+$ and $V_-$ as a function of the detuning for two cases,
depending on which of the potentials, open or closed, is deeper,
assuming they are both attractive.  In Fig.~\ref{structure1}(a) the
potential well in the open channel is deeper, and when we turn on
interactions, the dressed potential $V_-$ is shifted downwards from
$V_o$ and $V_+$ is shifted upwards from $V_c$.  Imagine that we fix
$V_{oc}$ (later we use $\tau$ instead) and increase $\nu$.  The
potential $V_-$ will converge back to $V_o$ as $\nu$ is increased, and
any bound states that may have be created at $\nu=0$ due to the
interaction that deepens the well will eventually be pushed out when
the detuning $\nu$ becomes large.  Consequently $V_+$ will eventually
approach $V_c+\nu$, and if the interaction pushes out bound states at
$\nu=0$ from the $V_c$ potential into continuum, they will eventually
come back at large $\nu$.  These observations are crucial for further
analysis, as we claim below that bound states appear whenever a FR
from the continuum is pushed under the dissociation threshold, and
similarly, FRs appear whenever a bound state from the continuum is
pushed out of the channel.

The situation is somewhat more complicated when the closed channel
potential is deeper than the open channel potential, as illustrated in
Fig~\ref{structure1}(b).  Now $V_-$ starts below $V_c$ for small
detuning and $V_+$ is above $V_o$ when the coupling is first turn on.
As we increase the detuning $\nu$, holding the coupling strength
fixed, we observe an avoided crossing: $V_-$ tends to $V_o$ and $V_+$
tends to $V_c+\nu$.  This case is carefully analyzed in conjunction with
our numerical studies reported in the next section.  The above
considerations allow us to determine the depth of the dressed potentials
and hence to find number of bound states in each of them for different
values of the strength and the detuning parameter $\nu$.

The number of $s$-wave bound states in the three-dimensional potential
well of depth $V$, $n$, is known to be equal to the value of the floor function for
$n^\star$, i.e., the largest natural number smaller or equal to $n^\star$,
which is defined by the formula:
\begin{equation}\label{depth}
    V = \frac{\hbar^2 }{2 m r_0^2} \left( \frac{\pi}{2} \right)^2
    (2 n^\star - 1)^2 .
\end{equation}
Note that $n^\star$ is in general not an integer. In the above formula
we deliberately did not use dimensionless parameters. {Consider the case where the weakest bound state is just below threshold, i.e., the potential $V$ is tuned such that there is a pole of the Green's function at $E = E_b \approx 0^-$, where $E_b$ is the binding energy of the bound state.  As $V$ is slowly decreased, at some value of the potential, $V_{\mathrm{th}}$, the pole of the Green's function crosses threshold an energy $E_r \approx 0^+$. For a scattering energy $E=E_r$ there is a zero energy resonance.  Before crossing threshold, i.e., for $V>V_{\mathrm{th}}$, the approximate formula for the scattering length is $a \approx \hbar/\sqrt{2m |E_b|}$; this formula is valid close to threshold} {\footnote{{ Note that the scattering length can be calculated for both $V>V_{\mathrm{th}}$ and for $V<V_{\mathrm{th}}$ in terms of the $s$-wave scattering phase shift $\delta(k)$, where $k=\sqrt{E}$. The desired relation for the scattering length is $a=-\lim_{k \to 0} \tan \delta(k)/k$. For $V>V_{\mathrm{th}}$, $\delta(0)=n \pi$ and $a=-[d \delta/dk]_{k \to 0}$, which is approximately equal to $\hbar/\sqrt{2m |E_b|}$. On the other hand, for the case of zero energy resonance ($V \to V_{\mathrm{th}}$ from below), $\delta(0)=(n+ \tfrac{1}{2}) \pi$ and, strictly speaking,  $a \to - \infty$. When the expression $- \tan \delta(k)/k$ is employed at finite but small $k$, $a$ is very large and negative.}}}.  Exactly at the point when one of the bound state crosses threshold, $n^\star$ becomes an integer and the number of bound states, including the one leaving the potential well, is precisely equal to $n=n^\star$.
Consequently, by equating the
potential depth in Eq.~\eqref{depth} to $V_\pm$ in Eq.~\eqref{veff}
and setting $\nu=0$, we can derive a formula for the threshold value
of the coupling strength $\tau_\mathrm{th}$, which constitutes the
boundary between regions in $\kappa_c$-$\kappa_o$ space where the
number of bound states increases or decreases by one. In dimensionless
units $\tau_\mathrm{th}$ is given by
\begin{equation}  \label{adiab}
    \tau_{\mathrm{th}}^2 = { \left[\left(\frac{\pi}{2}\right)^2
    (2 n-1)^2 - \kappa_o^2\right]\left[ \left(\frac{\pi}{2}\right)^2
    (2 n-1)^2 - \kappa_c^2\right]}.
\end{equation}

\begin{figure}[!ht]
\centering
\centering\subfigure[]{\includegraphics[width=0.45\textwidth]
{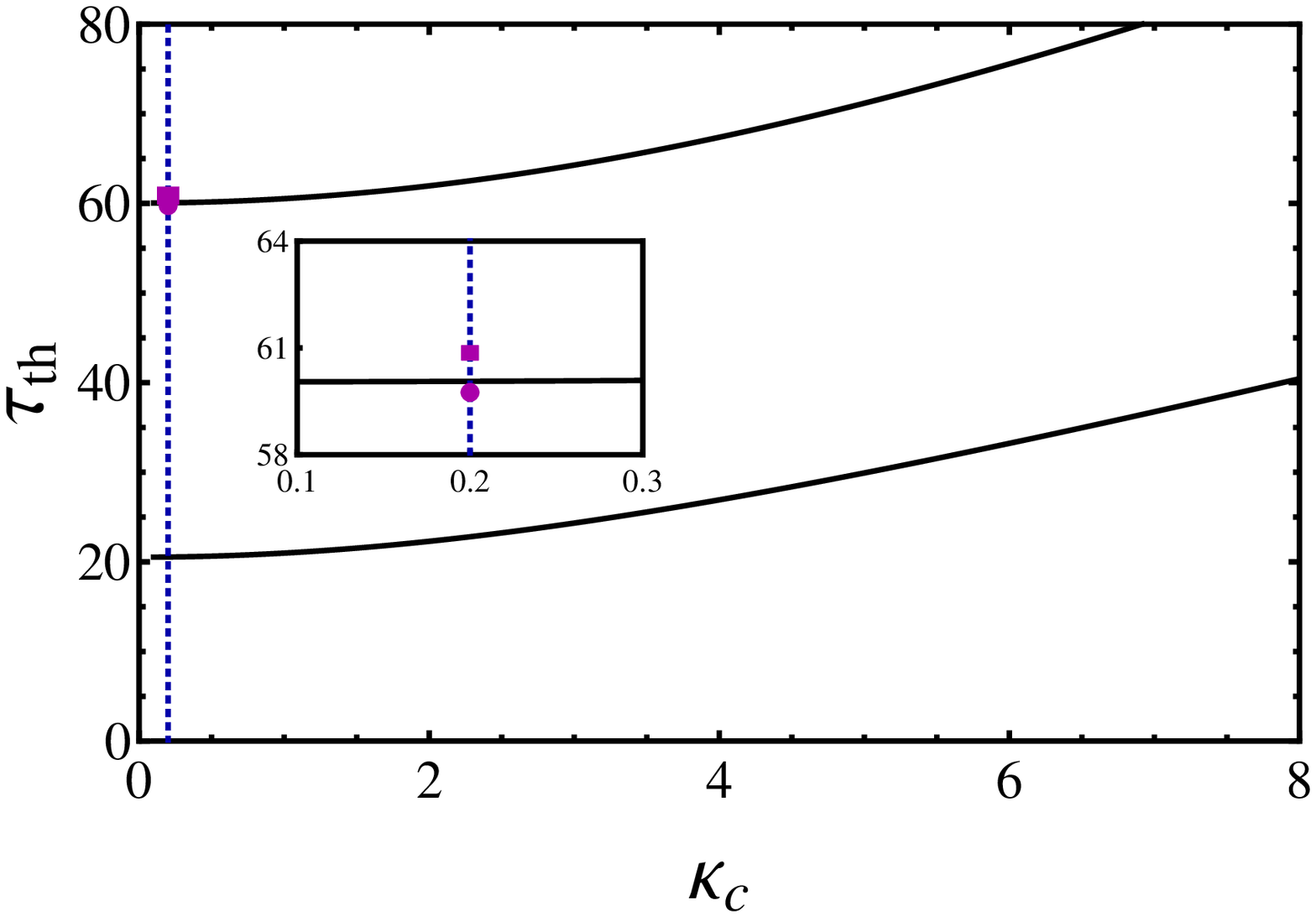}}
\centering\subfigure[]{\includegraphics[width=0.45\textwidth]
{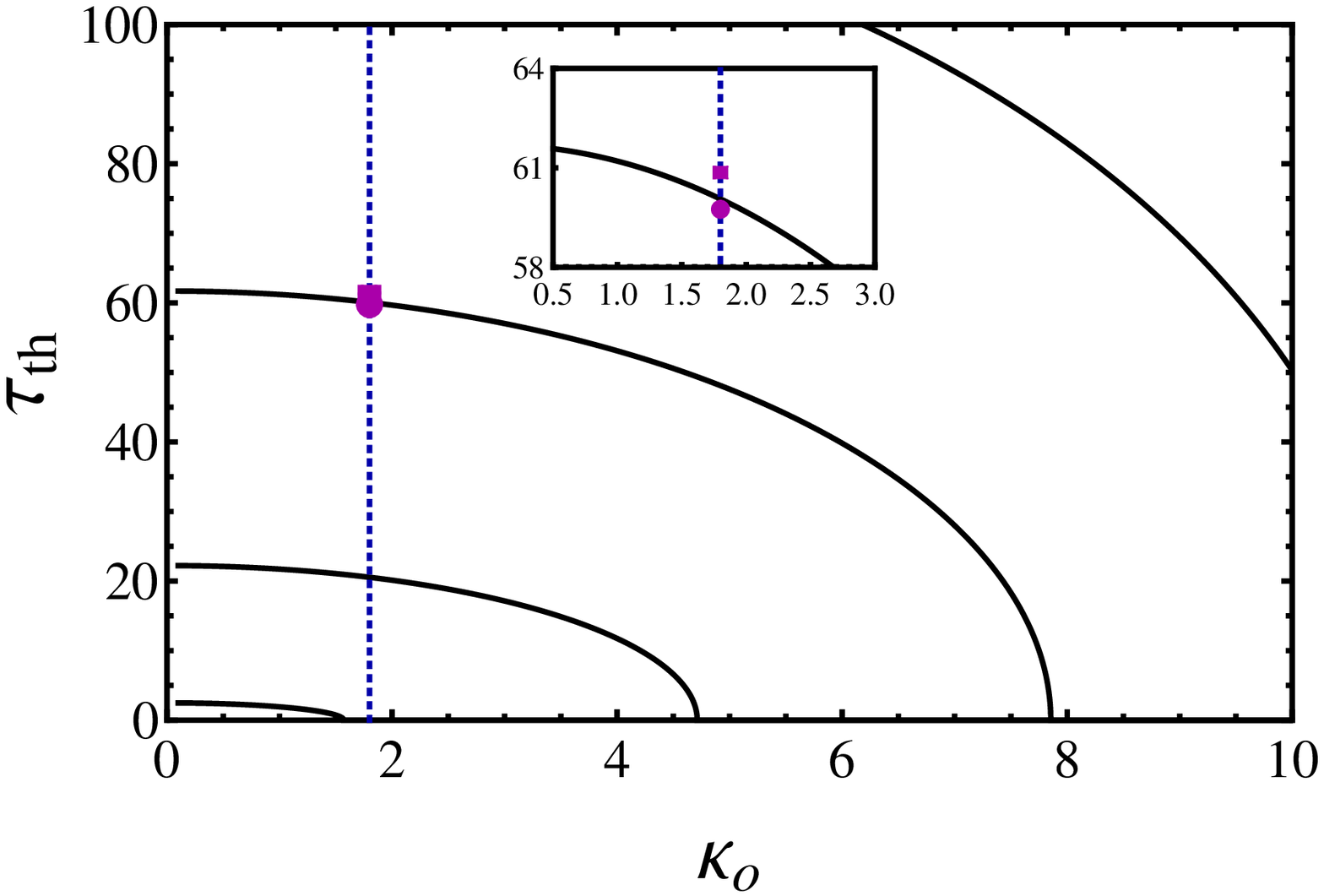}}
\caption{(Color online) Feshbach resonance for a
repulsive potential in the closed channel.  (a) Threshold curves (same
as in Fig.~\ref{atrakt1}) plotted versus the depth of the closed
channel (with $\kappa_o=1.8$).  (b) Threshold curves plotted versus
the depth of the open channel (for $\kappa^2_c=-0.04$).  The dashed
blue vertical line refers to the numerical calculations where we
keep the depth of the potential fixed ($\kappa_o=1.8$ and $\kappa^2_c=-0.04$)
and increase the value of $\tau$ in the analysis illustrated in
Figs.~\ref{repul2}-\ref{repul3}.  The purple circle (square) on this
line marks the value of $\tau=59.75$ ($\tau=60.86$) which is used in
red line in Fig.~\ref{repul2} (black lines in Figs.~\ref{repul2} and \ref{repul3}).  Because
the circle and square are almost coincide on the scale of the figure,
 we added an appropriate inset.}
\label{repul1}
\end{figure}

\begin{figure}[!ht]
\centering
\includegraphics[width=0.45\textwidth]{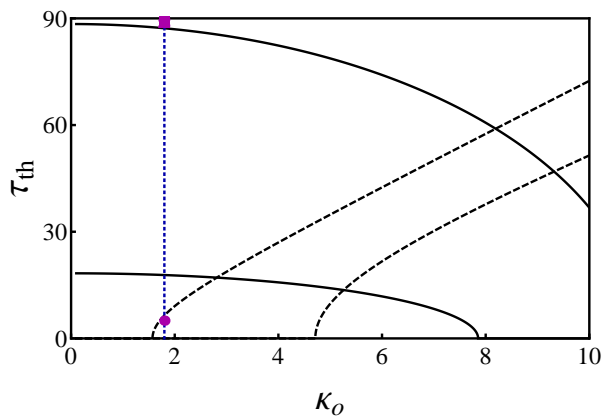}
\caption{(Color online) Feshbach resonances in the
case of an attractive closed channel potential.  The threshold values
of the coupling $\tau$ at which successive resonances appear are
plotted as a function of $\kappa_o$.  Each solid curve plotted versus
the depth of the closed channel potential (for fixed value of $\kappa_c=7.5$)
represents one such threshold.  When a solid curve is crossed, a new
Feshbach resonance emerges and when a black dashed curve is crossed,
one resonance disappears.  The dashed blue vertical line refers to our
numerical studies, as discussed in the text, where the depth of the
potentials is held fixed ($\kappa_o=1.8, \, \kappa_c=7.5$) and the
value of $\tau$ is increased.  The purple cirlce [square] on this line
marks the value of $\tau = 6.3$ ($\tau = 87.5$) which is used in
Figs.~\ref{atrakt2}(a) and \ref{atrakt3}(a) [Figs.~\ref{atrakt2}(b)
and \ref{atrakt3}(b)].}
\label{atrakt1}
\end{figure}

As seen from the above equation, there are two possibilities:
$\left(\frac{\pi}{2}\right)^2(2 n-1)^2 \geqslant \kappa_o^2$ and
$\left(\frac{\pi}{2}\right)^2(2 n-1)^2 \geqslant \kappa_c^2$ or
$\left(\frac{\pi}{2}\right)^2(2 n-1)^2 < \kappa_o^2$ and
$\left(\frac{\pi}{2}\right)^2(2 n-1)^2 < \kappa_c^2$.  The first
(second) describes the solid (dashed) lines on Fig.~\ref{atrakt1} and
depicts the emergence (vanishing) of the bound states in the lower
(higher) channel with the increasing coupling.  In the case of a
repulsive potential in the closed channel, there are no bound states,
but Eq.~\eqref{adiab} is still valid (cf., Fig.~\ref{repul1}).

Let us analyze the number of bound states in the dressed effective
potentials from the point of view of FRs.  The coupling of the
channels with strength $\tau$ and the shifting of the one channel
potential relative to the other by changing the asymptotic potential
energy difference $\nu$ can be considered as a two-step process
depicted in Fig.~\ref{fig:potentialadiab}~(a)-(c).  In the first step,
increase $\tau$ but keep $\nu=0$.  The deeper of the potential wells
becomes even deeper and a new bound state will eventually appear in
the adiabatic picture as $\tau$ is increased.  On the other hand, the
upper channel potential well becomes shallower and its bound states
(if present) can be pushed away (removed).  In the second step, for a
given $\tau$, we consider the process of increasing the asymptotic
potential energy difference between channels, $\nu$.  Every time a
bound state of the shallower channel crosses the asymptotic energy of
the open channel, we observe an upper channel dominated FR. With
increasing $\nu$, the coupling becomes less significant and the depths
of the channels gradually return to the case of $\tau=0$.  During this
process, bound states of the deeper channel disappear.  Every time a
bound state crosses the asymptotic potential energy of the open
channel, we observe a lower channel dominated resonance of the
scattering length.  Hence, the overall number of FRs, $n_\mathrm{FR}$,
for different $\nu$, is given by a simple formula,
\begin{equation}\label{NFR}
    n_\mathrm{FR} = n_{+} + (n_{-} - n_{o}),
\end{equation}
where $n_{+}$ is the number of the bound states in the upper
adiabatic potential for $\nu=0$, $n_{-}$ is the number of the
bound states in the lower adiabatic potential for $\nu=0$, and
$n_{o}$ is the number of the bound states in the open channel
potential.
{
\section{Strong-coupling regime}  \label{sec:nonpert}
}
The standard interpretation of FR phenomena applies for weak
inter-channel coupling strength $\tau$.  In this regime, a FR occurs
when the closed channel bound state energy approaches the incident
energy of the open channel.  But at sufficiently high values of the
coupling strength, there is a threshold above which yet another kind
of FR can appear.  This kind of FR occurs for strong coupling strength
even when the potential in the closed channel is {\em repulsive} (or
attractive but with no bound state present in the closed channel, as
this is possible in 3D).  We refer to this as the {strong coupling}
regime.

It was already shown that the tight-binding model reported in
Ref.~\cite{ABT_13} possess the strong coupling regime.  Here we
demonstrate using the adiabatic representation that the mechanism for
creation of this type of FR is similar to the original one, but this
time it is a state in a dressed potential that is approaching the
threshold rather than the initial bound state in the closed channel
potential.  We explain this mechanism in detail below.

There are significant differences in how the resonances behave,
depending on whether the closed channel potential is attractive, but
with no bound states in the closed channel, or repulsive.  Therefore,
we analyze the two cases separately, starting with the repulsive closed
channel potential case.

The expression for the scattering length in Eq.~(\ref{Eq:apha}) is
rather complicated, but the simple formula in Eq.~\eqref{adiab}
predicts the threshold of the coupling $\tau$, above which the new
FR appears.

\begin{figure}[!ht]
\centering
{\includegraphics[width=0.45\textwidth]
{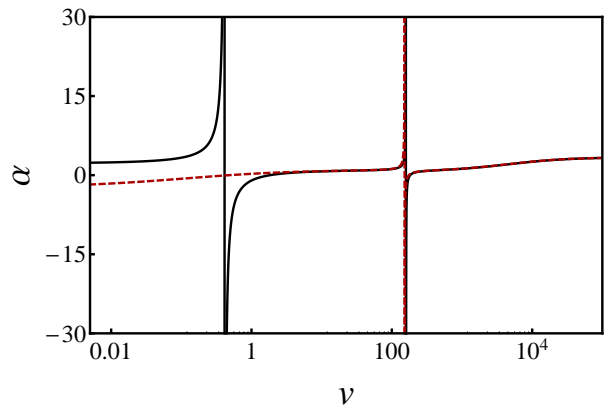}}
\caption{Scattering length in units of $r_0$ for a repulsive
potential in the closed channel as a function of $\nu$ (for
$\kappa_c^2=-0.04$).  Open channel supports one bound state
($\kappa_o=1.8$).  The dashed red curve shows the scattering length
for a coupling strength of $\tau = 59.75$ (purple circle in
Fig.~\ref{repul1}) which is below the second threshold at
$\tau_\mathrm{th} = 60.06$.  The solid black curve is for $\tau =
60.86$ (purple square in Fig.~\ref{repul1}) which is just above the
threshold.  See also Figs.~\ref{repul3}(a) and (b).}
\label{repul2}
\end{figure}

\begin{figure}[!ht]
\centering
{\includegraphics[width=0.45\textwidth]
{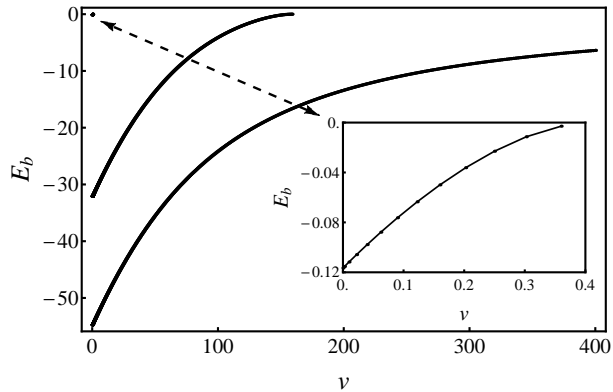}}
\caption{Energy of the bound states of the coupled system in units of
$\hbar^2/2m r_0^2$ for a repulsive potential in the closed channel
($\kappa_c^2=-0.04$) as a function of the energy shift $\nu$ (in the
same units as $E_b$).  The coupling strength $\tau=60.86$, is just above
threshold (purple square in Fig.~\ref{repul1}).  The inset shows the small
$\nu$ and $E_b$ region {and the dashed two-sided arrow depicts its position}.
}
\label{repul3}
\end{figure}

The following formula
\begin{equation}  \label{threp}
    \tau_{th}(n^\star) = \sqrt{\left(\frac{\pi}{2}\right)^2(2 n^\star
    - 1)^2 + |\kappa_c^2| } \, \sqrt{\left(\frac{\pi}{2}\right)^2(2
    n^\star - 1)^2 - \kappa_o^2},
\end{equation}
for the threshold value of the coupling strength $\tau$ for a repulsive
closed channel potential can be obtained from Eq.~(\ref{adiab}) simply by
changing $\kappa_c$ to $i \kappa_c$, which corresponds to the change from
attractive ($V_c \geqslant 0$) to repulsive ($V_c < 0$) closed channel
potential.

Our results for the case of a repulsive closed channel potential are
shown in Figs.~\ref{repul1} through \ref{repul3}.  Figure~\ref{repul1}
plots a set of solid curves marking the threshold for subsequent
resonances [see Eq.~(\ref{threp})] versus the depth of the closed
[Fig.~\ref{repul1}(a), $\kappa_o=1.8$] and the open channel
[Fig.~\ref{repul1}(b), $\kappa^2_c=-0.04$].  First we observe that, due
to the coupling, dressed potential picture emerges, and in particular
$V_-$ originates from open channel potential.  This potential gets
deeper when the interaction strength increases.  The exact depth of the
dressed potentials depends on both $\tau$ and $\nu$, but when we set
$\nu=0$ and increase $\tau$, a new bound state can appear in $V_-$.
For each fixed value of $\tau$ we study what happens to our system
with increasing value of the detuning $\nu$.  In the limit $\nu
\rightarrow \infty$, the depth of $V_-$ tends again to $V_o$ and
consequently a new bound state will move towards threshold,
eventually crossing it, thereby creating a FR.

In order to visualize this phenomenon we have drawn a dashed blue
vertical line in Figs.~\ref{repul1}(a) and (b), referring to the
numerical studies summarized in Figs.~\ref{repul2} and \ref{repul3}.
Moving vertically up along this line means keeping the depths of the
potentials ($\kappa_o=1.8, \, \kappa^2_c=-0.04$) fixed and increasing the
value of $\tau$.  Each time a solid curve is crossed, a new FR
emerges, due to the mechanism described in the previous paragraph.  In
Figs.~\ref{repul2} and \ref{repul3} we included two purple markers,
the circle and the square, for the values of $\tau=59.75$ and
$\tau=60.86$.  These values were used in Fig.~\ref{repul2}, where we
showed the scattering length as a function of $\nu$.  Notice that
there is only one FR for $\tau=59.75$, and two resonances for
$\tau=60.86$, in between these two values we crossed one threshold
line.

We can gain further insight regarding the mechanism described above
from Fig.~\ref{repul3}, where we plot energy of the bound states of
the coupled system as a function of the detuning $\nu$.  This figure
should be analyzed in parallel with Fig.~\ref{repul2}.  We notice that
FRs appear exactly when bound states of the coupled system disappear.
This is also the case when bound states of the potential $V_-$, induced
by coupling between channels, are pushed trough threshold into the continuum.

\begin{figure}[!ht]
\centering
\centering\subfigure[]{\includegraphics[scale=0.6]
{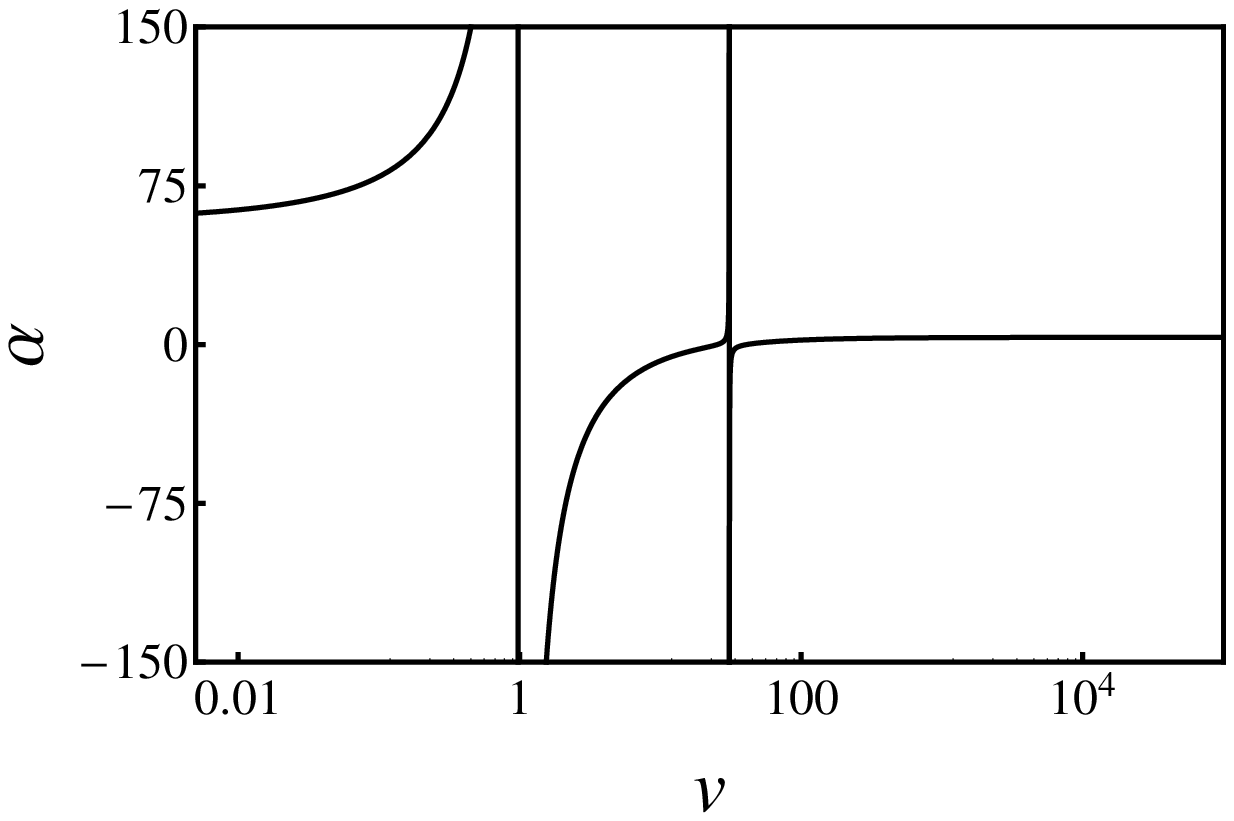}}
\centering\subfigure[]{\includegraphics[width=0.4\textwidth]
{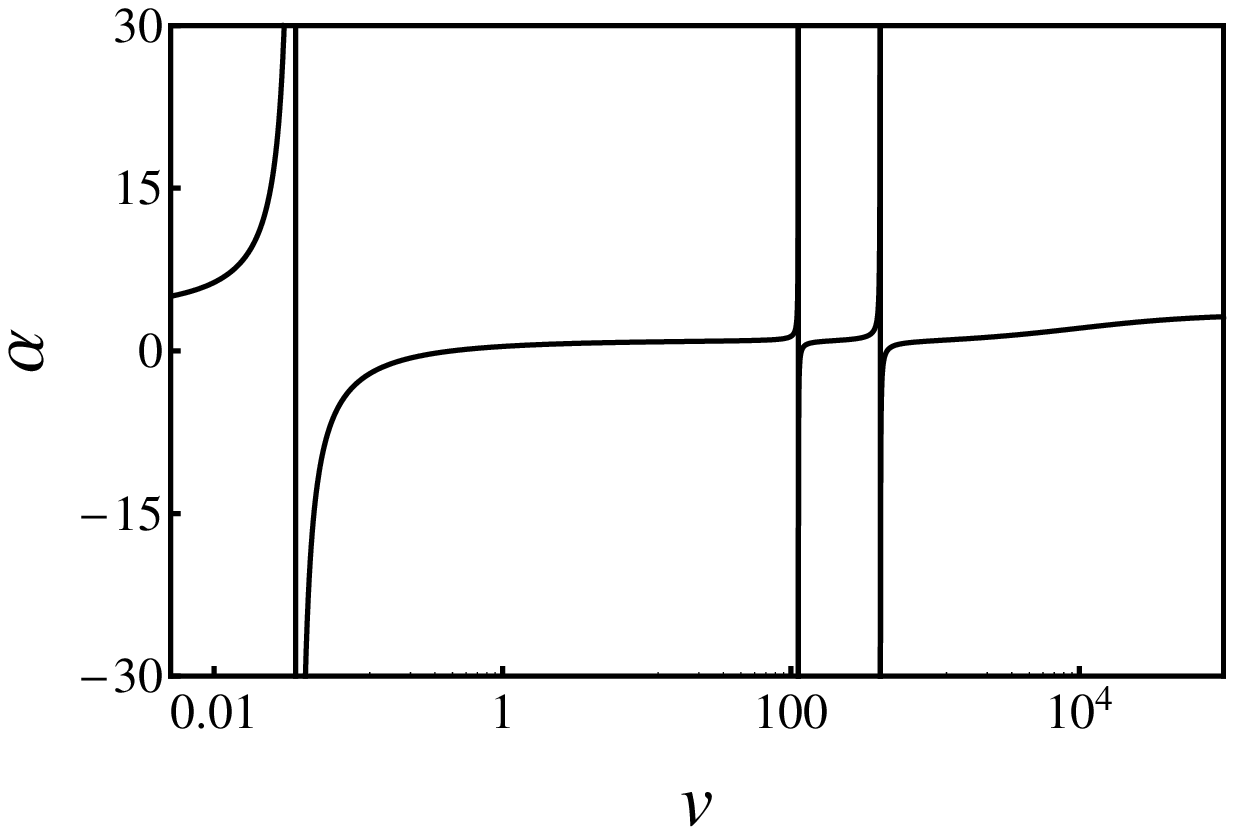}}
\caption{Scattering length in the units of $r_0$ as a function of
energy shift $\nu$ in units of $\hbar^2/(2mr_0^2)$ for an attractive closed channel potential with
$\kappa_c=7.5$.  Fixed value of $\kappa_o=1.8$ corresponds to the case
when open channel supports one bound state.  Coupling strength is
equal to (a) $\tau = 6.3$ (purple circle in Fig.~\ref{atrakt1}), and
(b) $\tau = 87.5$ (purple square in Fig.~\ref{atrakt1}).  Upon
increasing the values of coupling strength from (a) to (b) following
blue line (see Fig.~\ref{atrakt1}) one passes through three threshold
curves.  Two new resonances are created upon crossing the two solid lines
and one resonance disappears upon crossing the dashed line.
}
\label{atrakt2}
\end{figure}

\begin{figure}[!ht]
\centering
\centering\subfigure[]{\includegraphics[width=0.45\textwidth]
{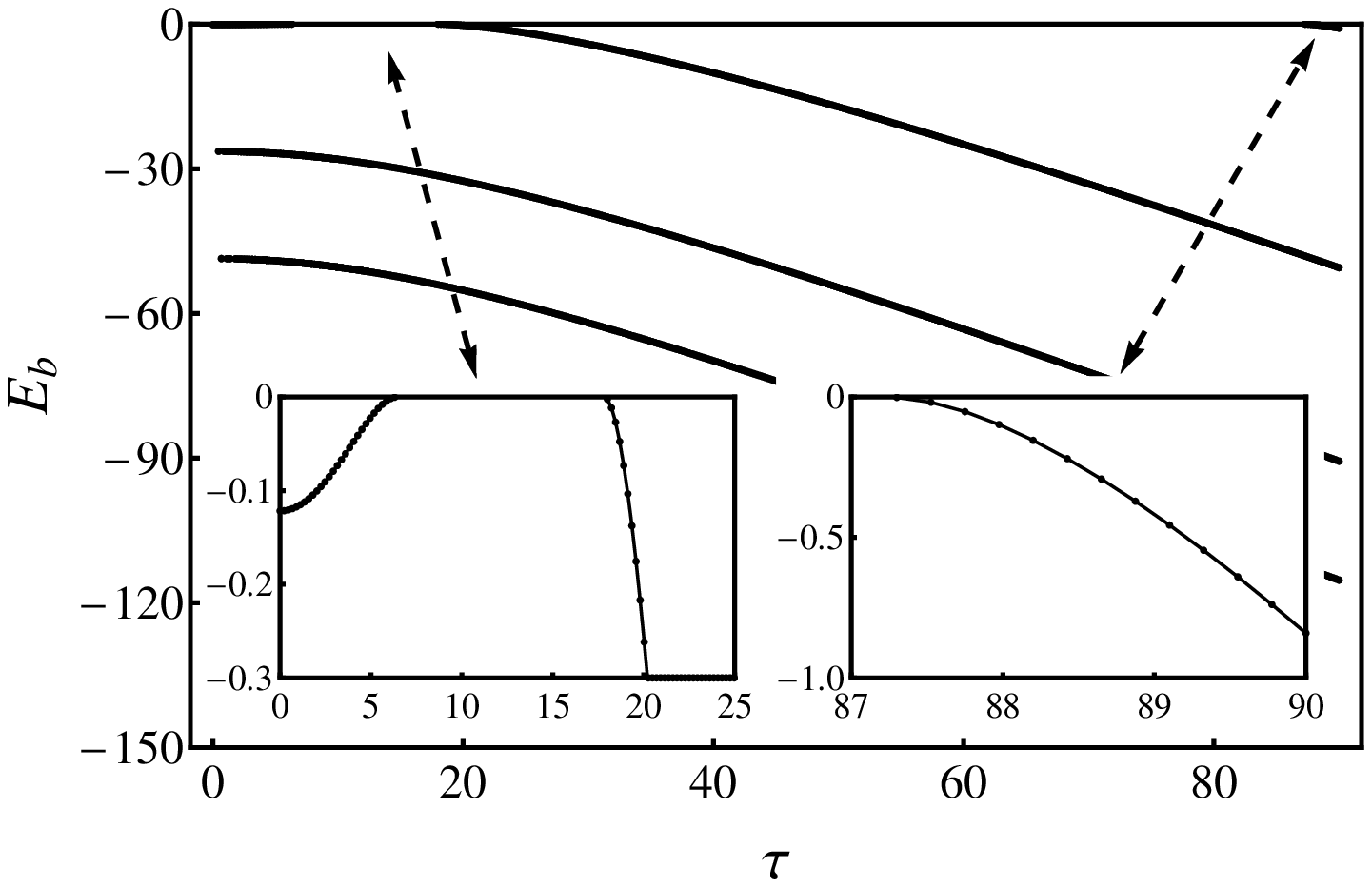}}
\centering\subfigure[]{\includegraphics[width=0.45\textwidth]
{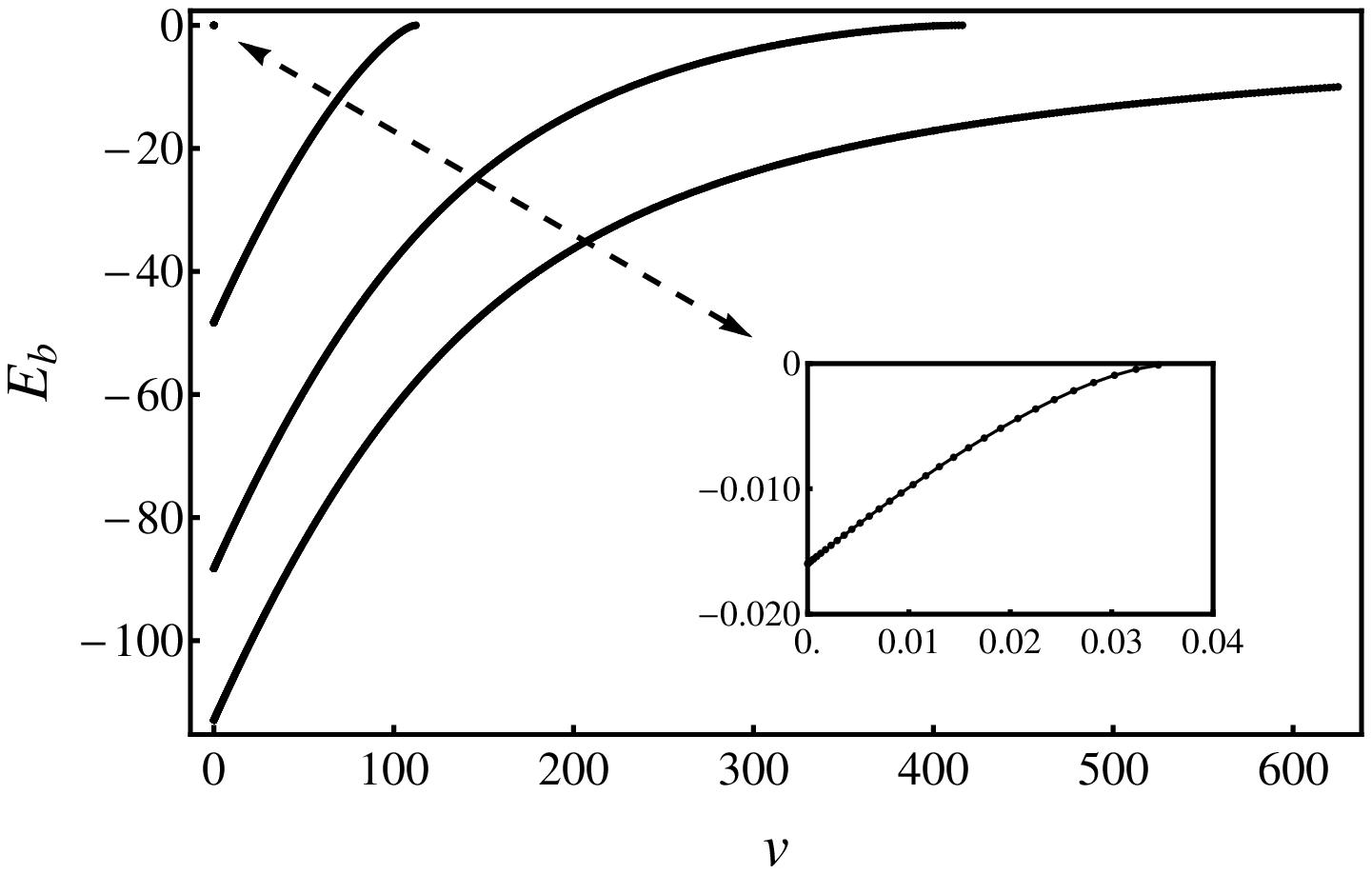}}
\caption{(a) Energy of the bound states of the coupled system
for an attractive potential in the closed
channel as a function of the coupling strength $\tau$ {for $\nu=0$}.  The values of
$\kappa_o$ and $\kappa_c$ are as in Fig.~\ref{atrakt2}.  The inset shows
the magnification of the top curve near the origin and reveals that
the bound state disappears through the threshold and returns.  (b)
Energy of the bound states for coupling strength $\tau=87.5$ (just
above the threshold value; corresponding to purple square in
Fig.~\ref{atrakt1}) as a function of the energy shift $\nu$.  The
inset shows the bound state energy for small values of $\nu$. All energies are given in units of $\hbar^2/(2mr_0^2)$.}
\label{atrakt3}
\end{figure}

The situation is slightly more complicated when there is an attractive potential
that supports a bound state in the closed channel, although, in general, a similar
mechanism applies.  The major difference between this and the previous
case is the existence of bound states in the closed channel.  The
detailed analysis is illustrated in Figs.~\ref{atrakt1} to \ref{atrakt3}.
Figure~\ref{atrakt1} displays the threshold coupling
$\tau$ at which resonances appear or vanish.  In Fig.~\ref{atrakt1}
the closed channel wavevector $\kappa_c$ is equal $7.5$.  When we fix
both $\kappa_o = 1.8$, and $\kappa_c = 7.5$ and increase the coupling
strength, we move along the vertical (blue) dashed line.  Taking this
path we obtain the results shown in Figs.~\ref{atrakt2} and
\ref{atrakt3}.  These figures show the scattering length versus
detuning at two particular values of coupling, $\tau=6.3$ and
$\tau=87.5$, which are marked with a circle and a square symbol in
Fig.~\ref{atrakt1}.  We observe that each time a solid curve is
crossed in the direction of increasing $\tau$, a new FR emerges, and
when a black dashed curve is crossed, one of the resonances
disappears.  Notice that two resonances are present for $\tau = 6.3$
in Fig.~\ref{atrakt2}(a) and the three resonances for $\tau = 87.5$ in
Fig.~\ref{atrakt2}(b).  As shown in Fig.~\ref{atrakt1}, between these
values of $\tau$, the blue line crosses three threshold curves, i.e.,
two solid curves and one black dashed curve.

In analogy with the repulsive case, let us now consider the bound
states of the coupled system and their dependence on $\nu$ and $\tau$.
Again we argue that FRs are associated with the position of the
coupled channel bound states.  Figure~\ref{atrakt3}(a) shows the
energy of the bound states versus the coupling strength $\tau$ {for $\nu=0$}.
Here we use the same values of $\kappa_o=1.8$ and $\kappa_c=7.5$ as in
Figure~\ref{atrakt2}.  We focus our attention on the threshold
($E_b=0$) and observe that as we increase the coupling strength
$\tau$, the highest bound state first disappears through the threshold
at $\tau=6.45$ and then comes back at $\tau=17.82$.  One can check
that the former event coincides with one bound state being pushed out
of the $V_+$ well and the former with a new bound state appearing in $V_-$.
At the right corner of Fig.~\ref{atrakt3}(a) we note the occurrence of
one more bound state of the coupled system, again associated with one
more bound state pushed into $V_-$.

When $\tau \neq 0$, it is advantageous to think of Feshbach processes
in terms of the adiabatic potentials $V_+$ and $V_-$.  Since in this
case the closed channel potential well is deeper than the open one, $V_-$
tends to $V_c$ as the coupling $\tau$ approaches zero.  For $\tau=0$ (no
coupling) the open channel dressed potential ($V_+$) supports one bound
state and the closed channel dressed potential ($V_-$) supports two bound
states.  For very small $\tau$ we expect two FRs as predicted by
Eq.~\ref{NFR}; they are shown in Fig.~\ref{atrakt2}.  But upon increasing
$\tau$ further we cross the first threshold value $\tau_\mathrm{th}$ shown
in Fig.~\ref{atrakt1}, and one of the resonances disappear.  It happens when
the bound state of the coupled system reaches threshold (see inset of
{Fig.~\ref{atrakt3}}), or equivalently, when a bound state in the dressed
potential $V_+$ is pushed into the continuum (remember that with increasing
$\tau$,  $V_-$ gets deeper and $V_+$ gets shallower). {As we increase the
coupling further, the next threshold is crossed and one additional bound
state appears in the $V_-$ potential and at the same time an additional
bound state appears in the coupled channel system {(see inset of Fig.~\ref{atrakt3})}.
If we increase $\tau$ even further, this pattern repeats; a new bound state
is formed (e.g., this occurs at $\tau=87.5$) in the $V_{-}$ potential and
in the coupled channel system.  In Fig.~\ref{atrakt2}, there are three FRs
and three bound states of the coupled system will approach zero when we
increase the detuning {as can be seen from Fig.~\ref{atrakt3}(b)}.}

\section{Summary and Conclusion}   \label{Sec:Summary}


The possibility of achieving Feshbach resonance without having a bound-state in the closed channel was suggested in our earlier work, based on a tight-binding model with very simple potentials \cite{ABT_13}. In the present work we expand upon that result and show that the concept is robust. Specifically, we studied a piecewise constant potential model and demonstrated that, for strong enough coupling strength, Feshbach resonance occurs even when the closed channel does not have a bound state or even when the closed channel potential is repulsive.  Our findings complement and substantially expand upon the analysis of the tight-binding model and the formal scattering theory results reported in Ref.~\cite{ABT_13}. We introduced a dressed potential picture (an adiabatic representation) and elucidated the physical mechanism behind this phenomenon. We derived analytical formulae for the number of Feshbach resonances as a function of the depth of the potentials.

\begin{acknowledgments}
We would like to acknowledge the Foundation for Polish Science International
Ph.D.~Projects Programme cofinanced by the EU European Regional
Development Fund.  This research was supported by grants from the
the National Science Center (DEC-2011/03/D/ST2/00200, DEC-2011/01/B/ST2/02030 and
DEC-2012/07/N/ST2/02879) and the Israel Science Foundation
(Nos.~295/2011 and 400/2012).
\end{acknowledgments}

\end{document}